\documentclass[preprint,pre,aps]{revtex4}
\usepackage[dvips]{graphicx}
\usepackage{color}
\bibstyle{prsty}
\begin{document}
\title{Effect of aggregation on adsorption phenomena }
\author{ M. Litniewski and A. Ciach}
\address{Institute of Physical Chemistry, Polish Academy of Sciences, 01-224 Warszawa, Poland}
\date{\today} 
\begin{abstract}
Adsorption at an attractive surface in a system with particles self-assembling into small clusters is studied by
Molecular dynamics (MD) simulation. We assume Lennard-Jones plus repulsive Yukawa tail interactions, and 
focus on small densities. The relative increase of the  temperature
at the critical cluster concentration near the attractive surface (CCCS)
shows a  power-law dependence on the strength of
the wall-particle attraction. At temperatures below the CCCS, the adsorbed layer consists of undeformed clusters
 if the wall-particle attraction is not too strong. 
Above the CCCS, or for strong
attraction leading to flattening of the  adsorbed aggregates, we obtain a 
monolayer that for strong or very strong attraction consists of flattened clusters or stripes respectively. 
The accumulated repulsion from the particles  adsorbed at the wall
leads to a repulsive barrier that slows down the adsorption process, and the accession time grows 
rapidly with the strength of the wall-particle attraction.
Beyond the adsorbed layer of particles, a  depletion region  of a thickness comparable with
the range of the repulsive tail of interactions occurs, and the density in this region decreases 
with increasing strength of the wall-particle attraction.
 At larger separations, the exponentially damped oscillations of density
agree  with theoretical predictions for self-assembling systems.
 Structural and thermal properties of the bulk are also determined.
 In particular, a new structural crossover associated
 with the maximum of the specific heat, and a double-peaked histogram of the cluster size distribution are observed.
\end{abstract}
\maketitle
\section{Introduction}
Theoretical studies and computer simulations show that competing interactions can lead to formation of spherical 
or elongated clusters, networks or layers of particles, and that these aggregates can form ordered periodic
patterns at low temperature
\cite{imperio:04:0,archer:07:1,archer:08:0,ciach:08:1,ciach:10:1,kowalczyk:11:0,zhuang:16:0,zhuang:16:1,edelmann:16:0,pini:17:0}. 
In experiment, three-dimensional (3D) periodic structures have not been detected yet, 
but the clusters and the network have been observed in a number of systems~\cite{campbell:05:0,stradner:04:0,royall:18:0,bergman:19:0}.
In these systems, the effective interactions between
particles are attractive at short distances and repulsive at large distances (SALR). 
The repulsion is often of electrostatic origin, and the attraction can be induced by the (complex) solvent. 
 Simulations of dilute systems reveal a structural crossover between individual particles (monomers) 
and clusters of a specific size. The borderline between the gas of 
particles at high temperature $T$ and small density $\rho$,
and the gas of clusters at low $T$ and large $\rho$, was termed critical
cluster concentration (CCC)~\cite{santos:17:0,hu:18:0}. It resembles 
the critical micelle concentration line in amphiphilic systems~\cite{amos:98:0,floriano:99:0},
and was defined in a similar way~\cite{santos:17:0}.

Adsorption phenomenon has been intensely studied for many systems, because of its significance for
various applications. 
To the best of our knowledge, however, the effect of aggregation and CCC on the adsorption and the 
near-surface structure has not been investigated yet. The two-dimensional models of particles 
interacting with the SALR potential
 show self-assembly into clusters, stripes and voids 
for increasing density~\cite{imperio:04:0,archer:08:0,almarza:14:0,pekalski:14:0}. 
These results can give some information about the structure of the first layer of particles adsorbed
at the surface.
However, in the case of aggregation, the gas contains clusters that are 3D
objects whose size and shape can be  changed.
When the attractive surface is in contact with such a gas, the adsorbed clusters can have different 
orientations and /or conformations, and different parts of them may occupy the first near-surface layer,  therefore  
 the 2D modeling 
 can be an oversimplification.
The questions whether the clusters are deformed near the surface, how  they are distributed
at different distances from the attractive wall, and how the aggregation influences the amount of adsorption for 
different strengths of wall-particle interactions are open. 

In this work we investigate the near-surface 
structure in a system 
consisting of spherical particles interacting with the Lennard-Jones potential plus repulsive Yukawa tail. 
We choose the potential leading to formation of small clusters. Small clusters were observed for example 
in Ref.\cite{campbell:05:0,stradner:04:0,kowalczyk:11:0,bergman:19:0}. 
We do not focus on a particular system, but on a generic model
that can reveal the general properties of adsorption in systems with competing interactions. We choose
molecular dynamics (MD) simulations and study the structural properties both in the bulk and near an 
attractive surface.

In sec.2, we define the model and describe briefly the simulation method. In sec.3, we present 
 the histograms for the cluster-size distribution, and determine the CCC temperature for three values of
density in the bulk. 
This temperature will serve as a reference for the CCC in a vicinity of the surfaces (CCCS) 
with different strengths of particle-wall 
attraction. The pair distribution function is also determined in this section, 
and compared with theoretical predictions of
the mesoscopic theory\cite{ciach:08:1,ciach:10:1}.
In sec.4, we determine the effect of the wall-particle interaction strength on the temperature at the CCCS for 
three values of density. In addition, we determine the density profiles for several strengths of
wall-particle interactions, and 
the adsorption in the near-surface layers of thickness $1.5\sigma$ and $2.6\sigma$, corresponding to a mono-
and bilayer of particles at the wall, respectively. In sec. 5 we summarize and discuss our results.

\section{The model and the method}

\begin{figure}[h]
 \centering
\includegraphics[scale=0.4]{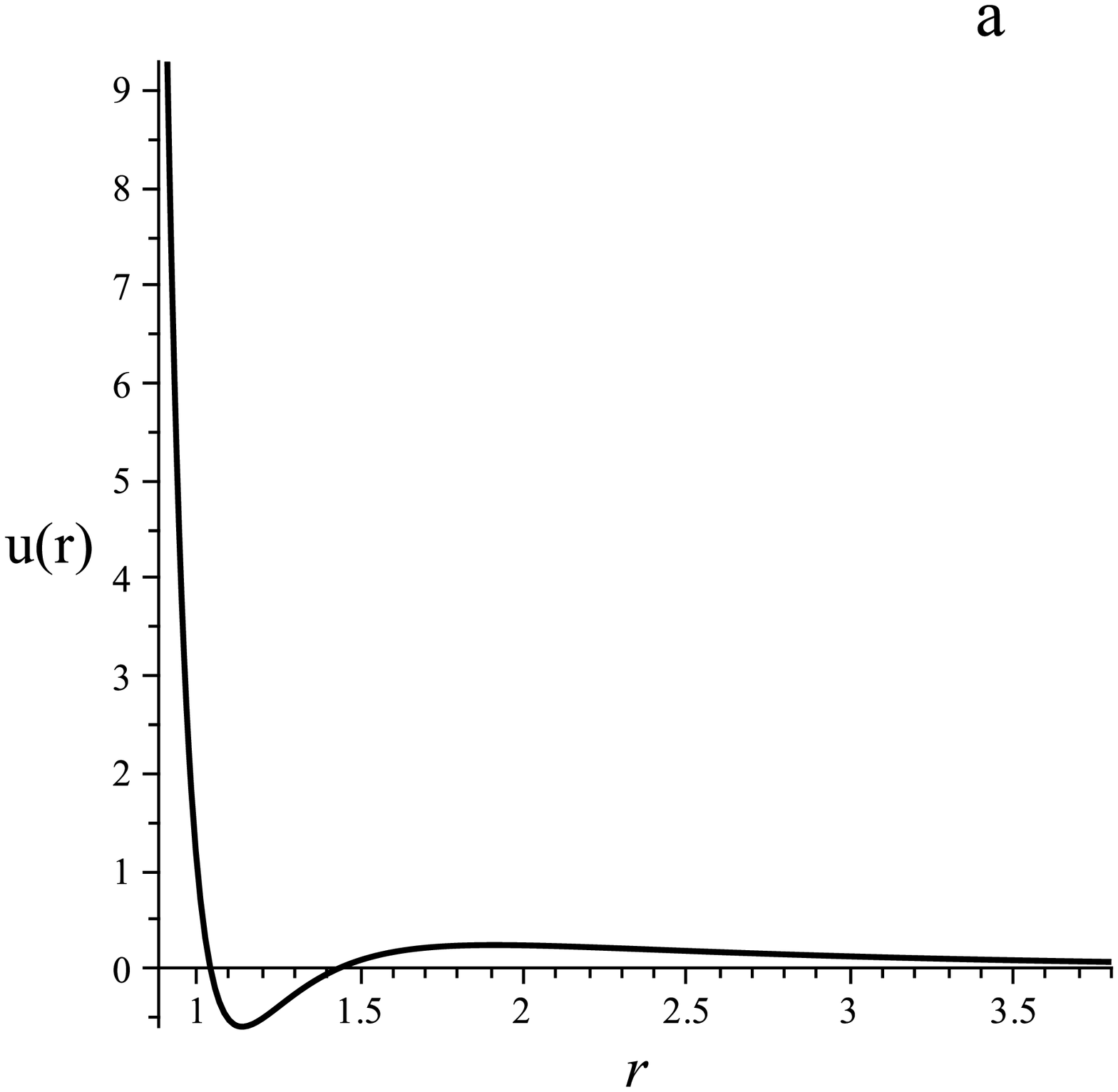}
\includegraphics[scale=0.4]{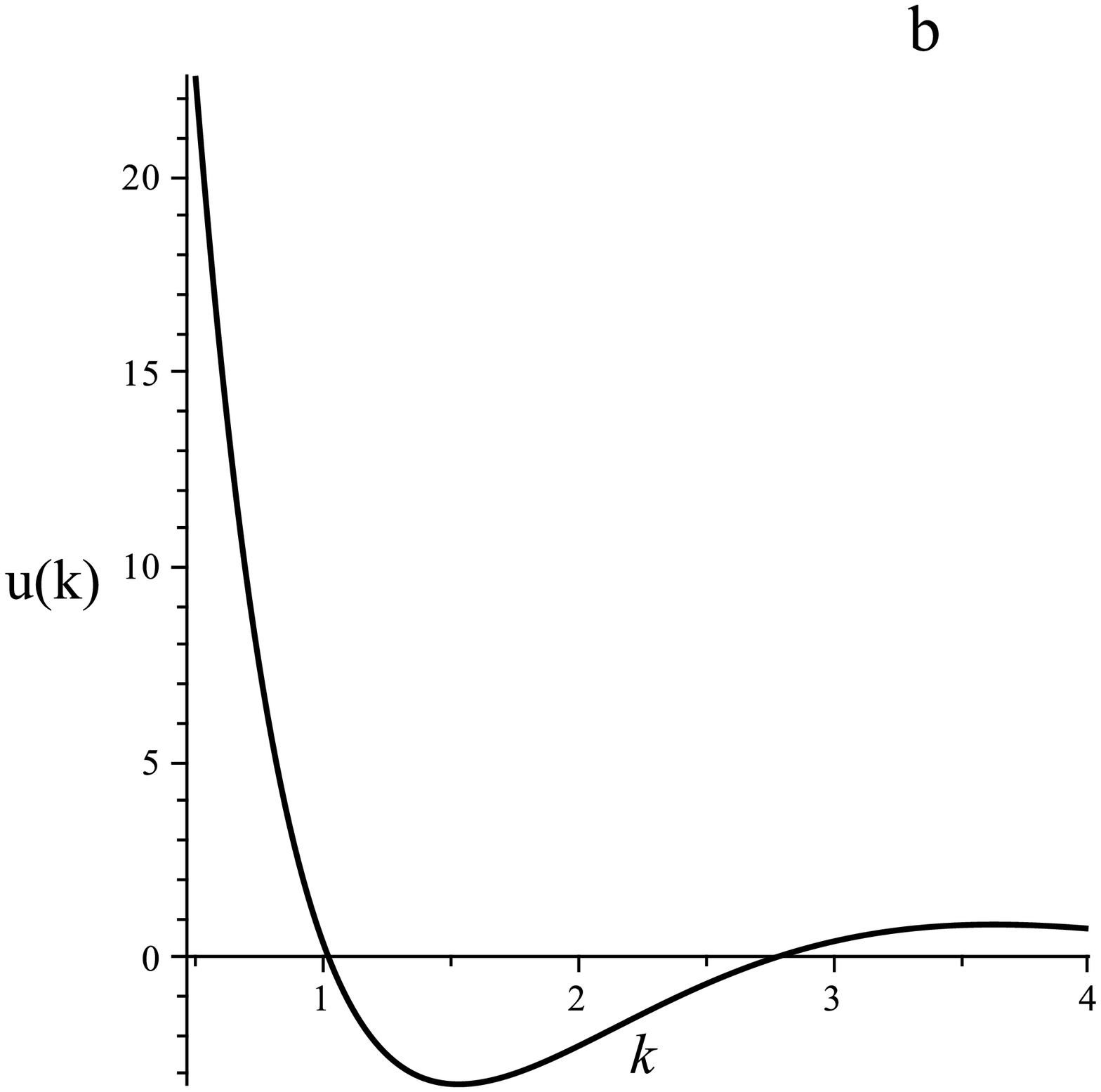}
\caption{The interaction potential $u(r)$ defined in (\ref{u(r)}) in the real space (a), and the
Fourier transform $\tilde u(k)$ of $\theta(r-1)u(r)$ (b). $r$ and $k$ are in $\sigma$ and $1/\sigma$ units,
respectively. {\color{black}The symbol indicates $r=r_{cs}$, the cutoff distance such that a particle separated by  $r<r_{cs}$ from its neighbor belongs to the same cluster.}}
\label{fig_u}
\end{figure}

{\color{black} For the interaction potential we choose} the sum of
 the short range Lennard-Jones (LJ) and the long range Yukawa potentials: 
\begin{equation}
\label{u(r)}
 u(r)=6\epsilon\Bigg[
 \Bigg(\frac{\sigma}{r}\Bigg)^{12} -\Bigg(\frac{\sigma}{r}\Bigg)^{6}
 \Bigg]+\frac{A}{r}e^{-r/\xi}
\end{equation}
where $\epsilon$ and $\sigma$ set the energy and length  units, and we assume $A=1.8$ and $\xi=2$. 
 The dimensionless temperature is defined 
as $T^*=k_BT/\epsilon$, where $k_B$ is  the Boltzmann constant. For clarity, the asterisk will be omitted.
The potential (\ref{u(r)}) is shown in Fig.\ref{fig_u}. {\color{black} The repulsion plays
an important role in the SALR systems, therefore in simulations the potential was truncated at a relatively 
large distance $r=6.75\sigma$, for which the interaction potential is very small ($u(r)<0.009$  for $r>6.75$).
We have performed additional test simulations for the cutoff $r=8$, and we got essentially the same results, 
with only a slight shift of the second maximum of the pair distribution function $g(r)$. 
}

To model the bulk, we consider $N=8000$ particles in a cubic box
with periodic boundary conditions in all three directions. 
The simulations were performed applying the constant temperature and
volume method~\cite{litniewski:93:0}. {\color{black}
We used this less known method because it is very simple and generates
canonical distribution in the coordinate space without using additional
variables and parameters (see also Ref.\cite{morishita:03:0,collins:10:0}). The method is easy to
implement by adopting the computer program used for classical constant energy
and volume simulations.} 
We focus on three values of the particle density,
$\rho=0.0005,0.0015,0.005$, and fix the
length $L$ of the edges of the cube accordingly.

In order to study the effect of the wall, the simulations were performed using the classical constant
energy and volume method~\cite{allen:90:0} for $N=12800$ particles in a rectangular box of the edges 
$L_x = L_y = L$ and $L_z \approx 1.6L$. The periodic boundary conditions are applied only along 
the $x$ and $y$ directions.
The $z$-th direction is restricted by two walls. For the left wall we assume the wall-particle interaction
of the form    
\begin{equation}
\label{VL}
 V_L(z)=4\gamma\epsilon\Bigg[
 \Bigg(\frac{\sigma}{z-z_L}\Bigg)^{12} -\Bigg(\frac{\sigma}{z-z_L}\Bigg)^{6}
 \Bigg]
\end{equation}
and for the right wall we assume only the repulsive interactions
\begin{equation}
 V_R(z)=0.5\epsilon
\Bigg(\frac{\sigma}{z_R-z}\Bigg)^{12},
\end{equation}
where $z_R-z_L=1.6L$. 
We shall investigate the structure near the left wall for the range of attractive interaction 
$0.5<\gamma<3$. 

In order to study the aggregation, we first introduce the distance criterion for particles forming the cluster.
For the SALR potential, the distance $r=r_{cs}$, where the pair potential (\ref{u(r)})
crosses zero {\color{black}for the second time (see Fig.1a)}
is a natural distance such that a particle separated by $r<r_{cs}$ from a particle belonging to a cluster, 
 belongs to the same cluster. The cluster size distribution is defined in the some way as in Ref.\cite{santos:17:0},
\begin{equation}
\label{p(M)}
 p(M)=\frac{MP(M)}{\sum_MMP(M)},
\end{equation}
where $P(M)$ is the probability of finding an aggregate of size $M$. 

{\color{black}The CCC is a structural crossover,
and cannot be defined in a  unique way. We adopt the criterion that the borderline between the monomers and 
the clusters is given by the inflection point at the $p(M)$ line defined in} (\ref{p(M)}). 

 {\color{black} In contrast to thermodynamic phase transitions, the CCC is a structural crossover, and the CCC line is not uniquely defined. 
 In fact the CCC is a crossover region rather than a line. We choose  the inflection point of $p(M)$ defined in (\ref{p(M)}), because it is a well-defined borderline between the histograms with one maximum (for monomers) and two maxima (for the monomers and the optimal clusters). A maximum of the histogram for a particular size of the cluster indicates that there exists an optimal size, thus signaling formation of a well-defined structure rather than random inhomogeneities in the system. Different criteria for the CCC would lead only to some small shifts of our line within the crossover region.}

\section{results for the bulk}

\subsection{the clusters}

 Simulation snapshots and the histograms (\ref{p(M)}) show formation of small clusters consisting of a few particles.
For such a discrete system, we determine the temperature at the CCC from the first appearance of $p(M)/p(M-1)>1$
when the temperature is decreased. In our case it happens for $M=4$ for the three considered densities. 
Analyzing $p(4)/p(3)$ as a function of $T$ for $\rho=0.0005, 0.0015, 0.005$, we estimated 
$T_{CC}$ (here, the moment that $p(4)/p(3) = 1$) as: $T=0.130, 0.143, 0.164$ for $\rho =  0.0005, 0.0015, 0.005$
respectively. 

\begin{figure}[h]
 \centering
 \includegraphics[scale=0.55]{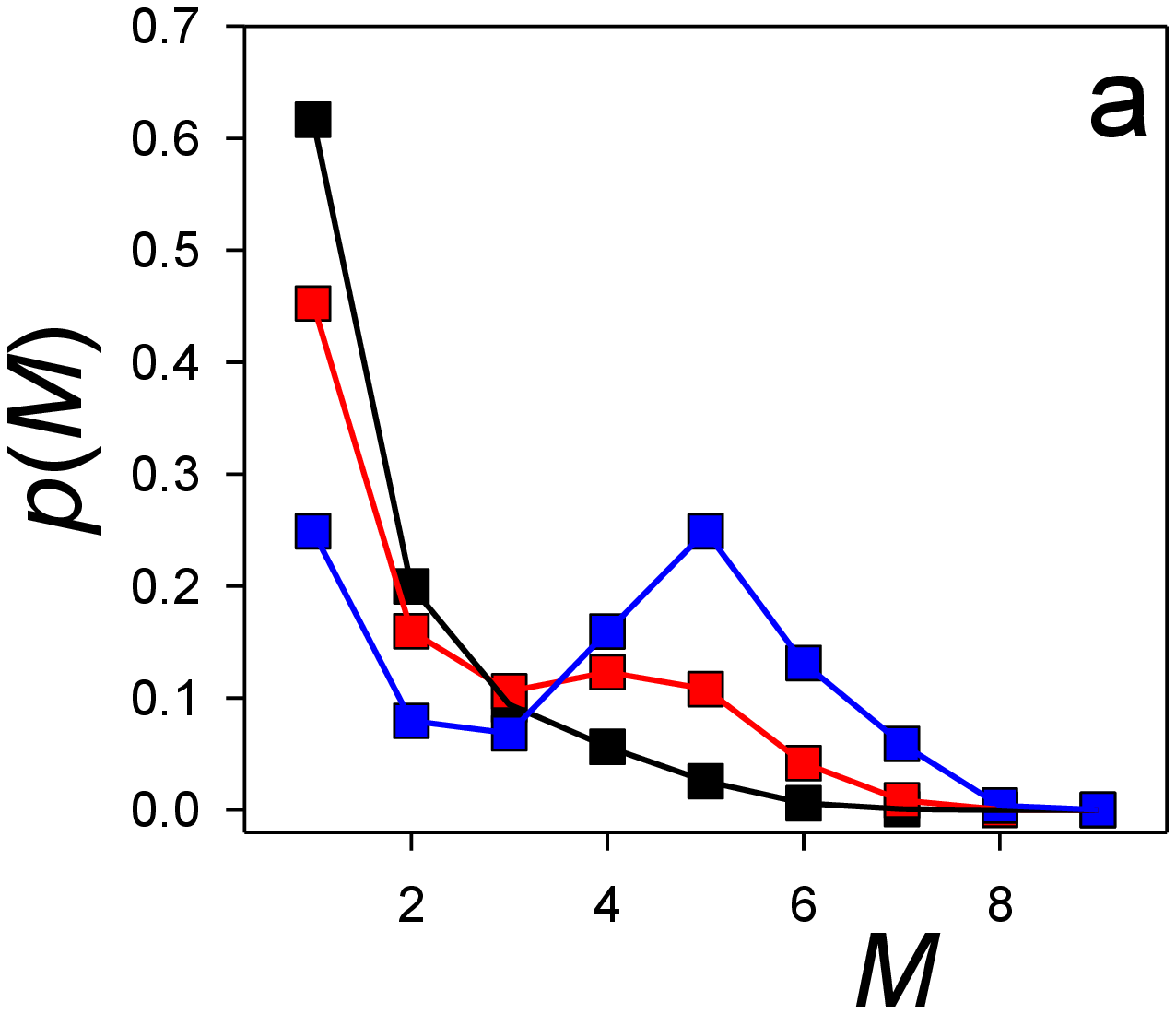}
  \includegraphics[scale=0.55]{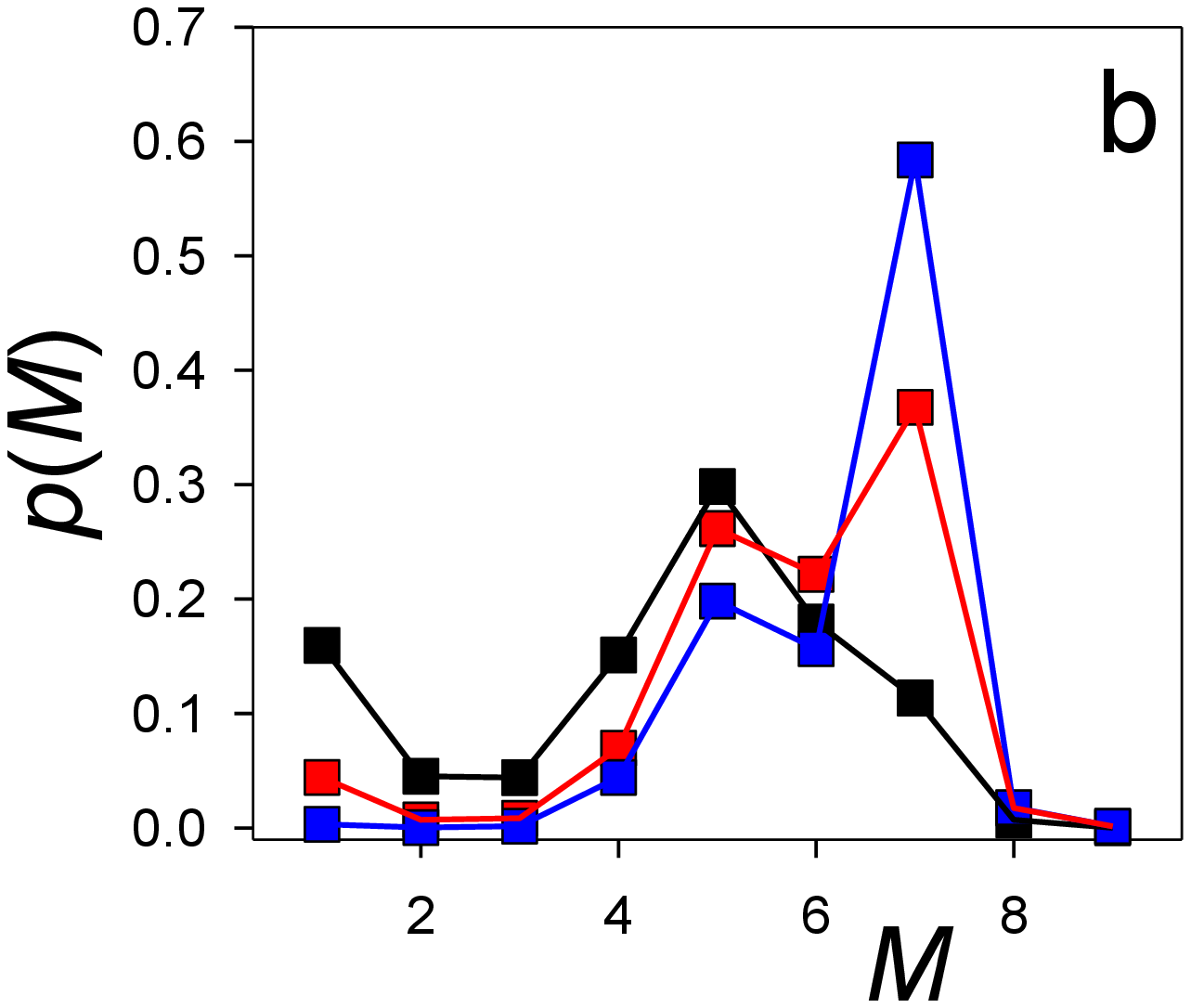}
\caption{Histograms for the probability of finding a particle in a cluster consisting of $M$ particles
for $\rho=0.005$. From the top to the bottom line on the left (black, red, blue), $T=0.18, 0.16, 0.14$ and  $T=0.13, 0.11, 0.09 $ 
on the (a) and (b) panels respectively. Lines are to guide the eye. }
\label{fig_his}
\end{figure}

The histograms for $\rho=0.005$ and $0.09<T<0.18$ are shown in Fig.\ref{fig_his}. 
At high $T$, $p(M)$ decreases monotonically, and for $T<0.164$ a maximum at $M=5$ appears.
Further decrease of $T$ below $T=0.12$  leads to a second maximum for $M=7$. {\color{black}Clusters composed of
more than $9$ particles are not formed, i.e. the distances between the particles within the cluster are smaller than the cutoff of the interaction potential, so
our analysis is not biased by the choice of the cutoff}.  

\begin{figure}[h]
 \centering
\includegraphics[scale=0.3]{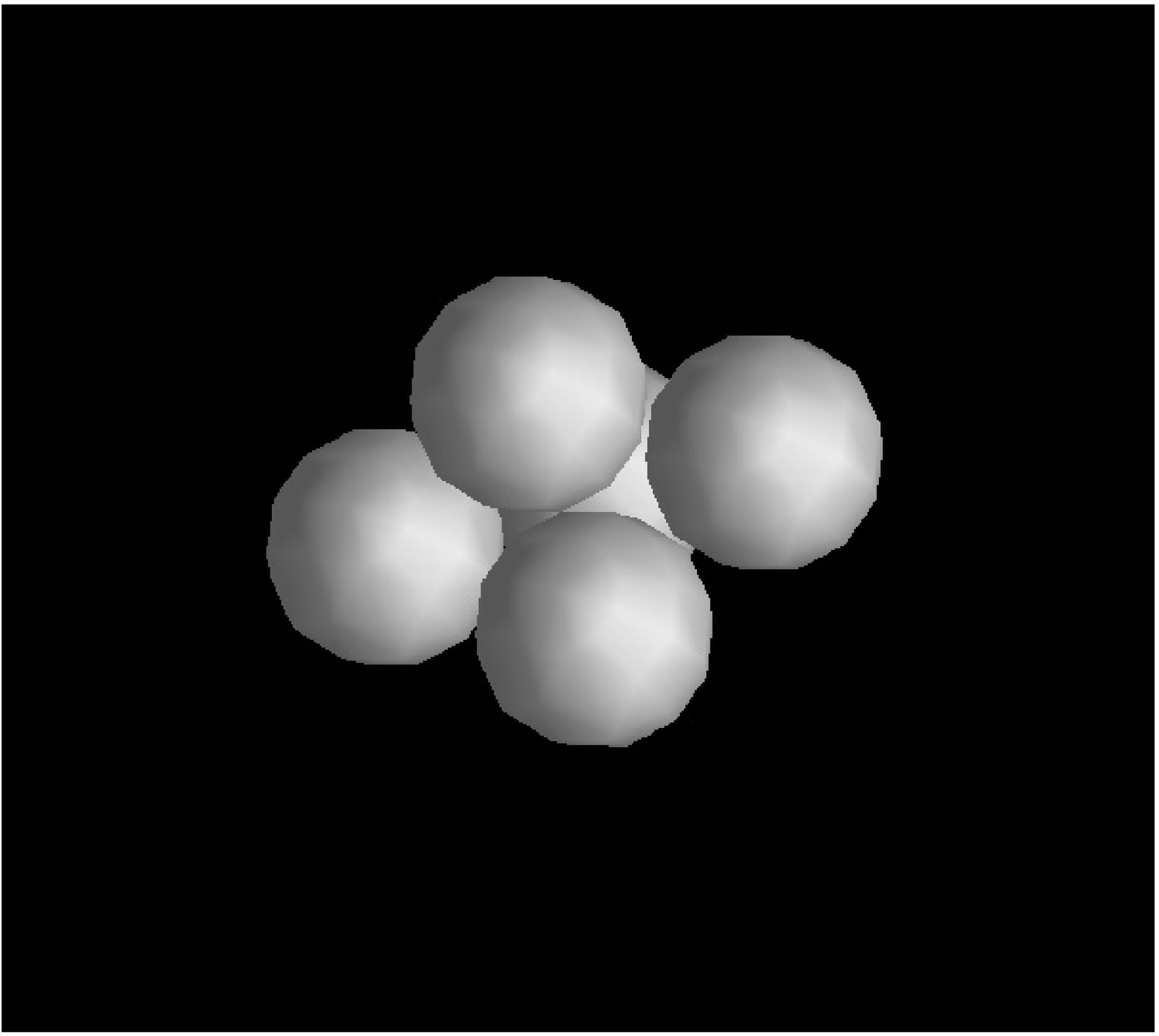}
\includegraphics[scale=0.3]{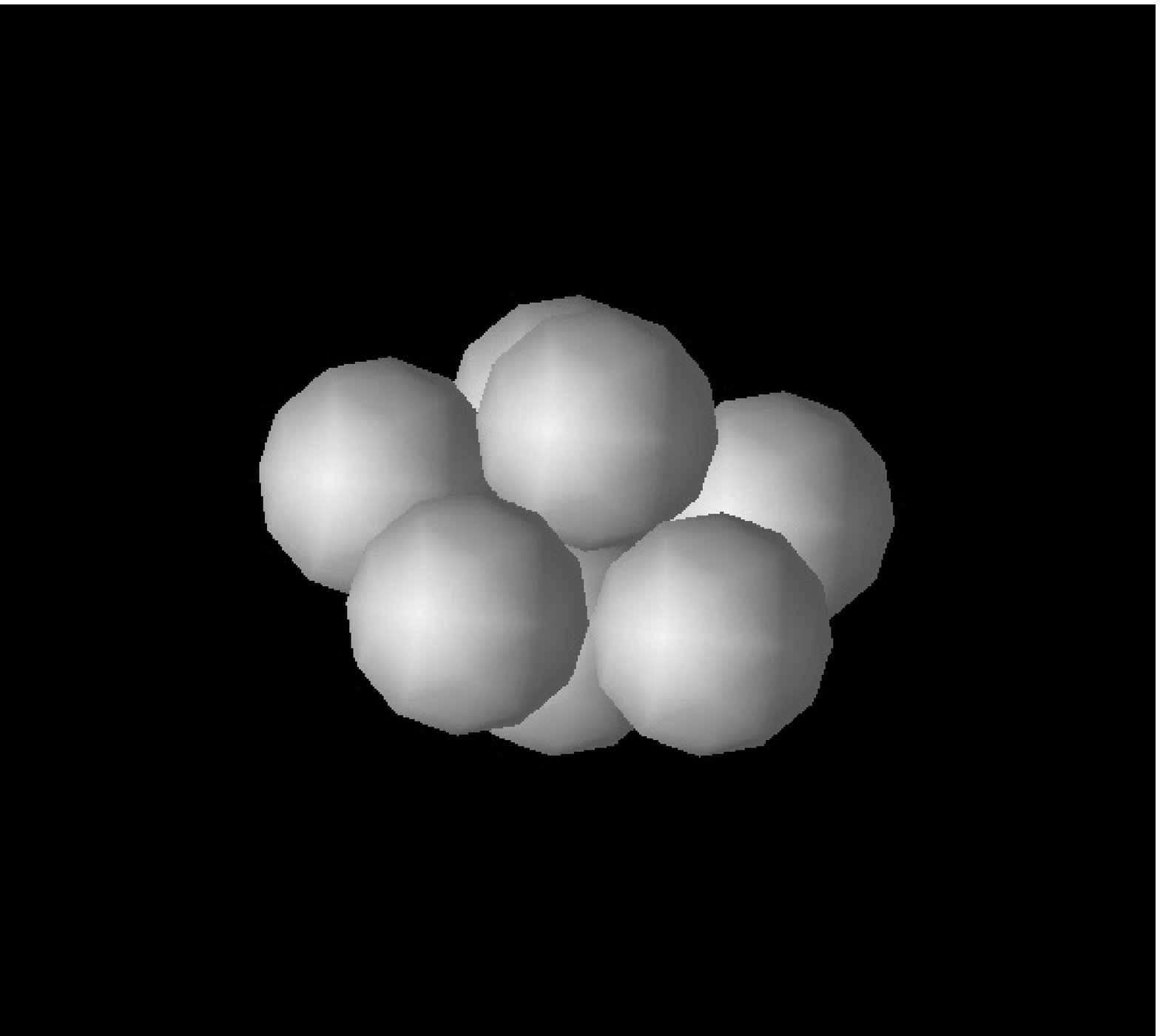}
\caption{Cartoon showing the clusters composed of 5  and 7  particles,
corresponding to the maxima of the histograms. }
\label{fig_clus}
\end{figure}

{\color{black}It is interesting to find the energetically favorable structure of the clusters. The structures corresponding to the minimum energy per particle should form for very low T.} In order to determine the structure of the clusters for $T\to 0$,
we decreased the temperature of the system from $T = 0.11$ (red line in Fig. 2b) to $T = 0.0025$. 
From the distribution of the interparticle distances we find
that in the 5 and 7 particle clusters, shown in Fig.\ref{fig_clus} ,
 3 and 5 particles form vertices of a regular triangle and pentagon with the edge length $a$, respectively.
 The two remaining particles 
are located above and below the center of the polygon at the line perpendicular to the  polygon plane. 
They are separated by $2a\sqrt{2/3}$ and $a$ for the 5 and 7 particle cluster respectively. 

For $M=5$, there is 9 pairs separated by the distance $a\approx r_{min}= 1.139$, and one pair 
of particles separated by 
$2a\sqrt{2/3}$. The energy per particle is $U(5)/5\approx -1.03$.
For $M=7$,  $a\approx 1.007r_{min}$, and there are 6 pairs separated by $a$, 10 pairs separated by $~0.99a$
and 5 pairs separated by $~1.62 a$, which gives  the lowest energy per particle, $U(7)/7\approx -1.21$.

{\color{black}The instantaneous thermal structures differ from the ones corresponding to the minimum of the energy due to the thermal motion and entropic effects that depend strongly on temperature. In particular, deformations of the bigger, 7-fold clusters due to thermal motion of particles may lead to instantaneous configurations corresponding to the repulsion of some particles from the cluster, and to formation of smaller clusters, as we indeed observe. Our results indicate that the 5-fold clusters are more stable against the thermal motion, despite higher energy per particle in the optimal  configuration.}
Only at low $T$, i.e. when the thermal motion is suppressed,
the dominating  clusters correspond to the lowest energy per particle, and
for $T\to 0$ only the 7-particle clusters remain. 

{\color{black}The fact that the 5-fold clusters dominate at high $T$ even though the  7-fold clusters have a lower energy in their optimal configuration is somewhat surprising. From the thermodynamic point of view, the reason is a fine balance between the energy $U$ and the entropy $S$ in the Helmholtz free energy $F=U-TS$. When the 7-fold clusters are formed, the number of objects (particles + clusters) decreases more strongly and it leads to a larger decrease of the entropy $S$ compared to the formation of the 5-fold clusters. At relatively high $T$ this effect may dominate over the increase of the energy (that is not large, especially for clusters deformed by thermal motion), therefore the 5-fold clusters may appear with higher probability than the 7-fold ones.  
}

\subsection{the pair distribution function}

\begin{figure}[h]
 \centering
 \includegraphics[scale=0.65]{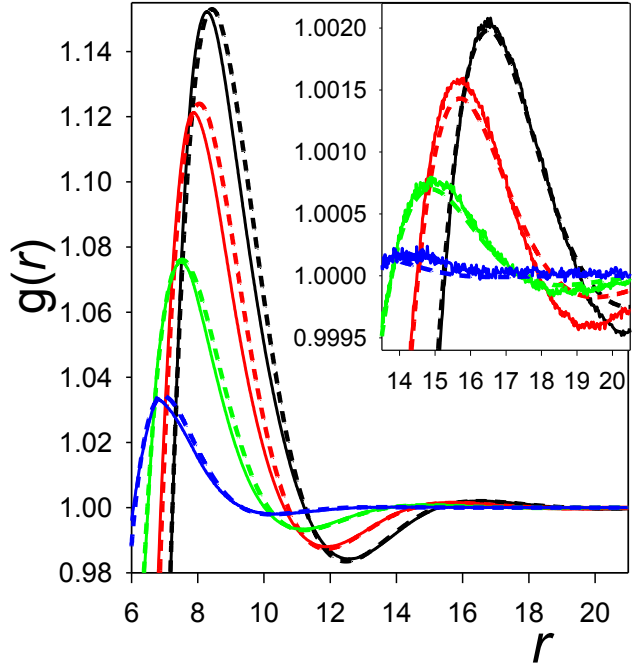}
\caption{Pair distribution function $g(r)$ for $\rho=0.005$ and $T=0.11$ (black line),
$T=0.13$ (red line), $T=0.15$ (green line) and $T=0.18$ (blue line). The dotted lines are the
simulation results, and the continuous lines are
the best fits to Eq.(\ref{gr}).  }
\label{fig_g(r)}
\end{figure}

The pair distribution function $g(r)$ for $\rho=0.005$ and temperatures $T=0.11, 0.13, 0.15$ and $T=0.18$ is shown
in Fig.\ref{fig_g(r)} for large separations $r>6$.
The solid lines in Fig.\ref{fig_g(r)} were obtained by fitting the simulation results to
the formula~\cite{ciach:08:1,ciach:10:1} that in general should be obeyed asymptotically for $r\to \infty$,
\begin{equation}
\label{gr}
 g(r)=1+\frac{A_0}{r}\sin(\alpha_1 r + \phi)e^{-\alpha_0r}.
\end{equation}
The fitting parameters are given in table 1.
\begin{table}
\begin{tabular}{|c|c|c|c|c|c|c|}
 \hline
 $T$&$\alpha_1^{f}$&$\alpha_0^{f}$&$\alpha_1^{t}$&$\alpha_0^{t}$&$\bar M$&$2\pi/\ell$\\
 \hline
 0.11&0.785&0.46&0.864&0.468&4.68&0.64\\
  \hline
  0.13&0.827&0.50&0.849&0.535&2.94&0.75\\
  \hline
  0.15& 0.861&0.55&0.820&0.638& 1.88&0.871\\
   \hline
   0.18& 0.971& 0.75&0.764&0.792&1.30&0.984\\
    \hline
\end{tabular}
\caption{Fitting parameters $\alpha_0^{f}$ and $\alpha_1^{f}$ for $g(r)$ approximated by Eq.(\ref{gr})
and shown in Fig.\ref{fig_g(r)}.
The parameters $\alpha_1^{t}$, $\alpha_0^{t}$ are
obtained from the mesoscopic theory,  Eq.(\ref{G-1}), and $\bar M$ is the average number 
of particles per cluster obtained form the histograms. In the last column, $2\pi/\ell$, where
 $\ell$ is the average distance between the objects ($\ell^3=\bar M/\rho$) is shown.
$T$ denotes temperature.}
\end{table}
Eq.(\ref{gr}) fits the simulation results reasonably well for $r>6$, i.e.
beyond the first period of the oscillatory decay. {\color{black} The  satisfactory agreement between 
our numerical results for $g(r)$ and the analytical expression (\ref{gr}) that is valid only for distances larger than the range of the interactions confirms again that cutting off the interactions for $r=6.75$ is justified}.

The fitting parameters can be compared with theoretical predictions of the mesoscopic density
functional theory (DFT).
In this mean-field (MF) theory~\cite{ciach:08:1,ciach:10:1}, the  Fourier transform $\tilde G(k)$
of  $G(r)=(g(r)-1)\rho^2$ is approximated by
\begin{equation}
 \tilde G(k)^{-1}=\beta\tilde u(k) -\frac{\partial^2s/k_B}{\partial \rho^2},
\end{equation}
 where $\beta=1/(k_BT)$, $s$ denotes the entropy per unit volume, 
 and $\tilde u(k)$
 denotes the Fourier transform of $u(r)\theta(r-1)$, 
 where $u(r)$ is the
 interaction potential (Eq.(\ref{u(r)})), and $\theta$ is the unit step function.
 With this definition of $\tilde u(k)$, we do not take into account contributions to the
internal energy from overlapping cores of the particles. 
For the considered dilute systems, the perfect gas approximation, $s/k_B=-\rho(\ln(\rho)-1)$, is 
 sufficiently accurate for  homogeneous gases. 
When some fraction of the particles aggregates into clusters, however, then  
 the number of objects - isolated particles plus clusters - is smaller than the number of particles. 
 When the number of objects in the dilute system decreases,
 the entropy decreases as well.
 In a very crude approximation, we can consider a perfect gas of $N/\bar M$ objects, where $\bar M$ is the 
 average number of particles per cluster. 
In this approximation, 
\begin{equation}
\label{G-1}
 \tilde G(k)^{-1}=\beta\tilde u(k) +\frac{1}{\bar M \rho}.
\end{equation}
As shown in the histograms and table 1, $\bar M$ depends on temperature. 

For the potential (\ref{u(r)}), $\tilde G(k)^{-1}$ takes a minimum for $k=k_0>0$ (Fig\ref{fig_u}). 
In such a case, in the real space representation  we obtain
the approximation, valid for large separations~\cite{ciach:08:1,ciach:10:1}
\begin{equation}
 G(r)=\frac{A_0\rho^2}{r} \sin(\alpha_1 r+\phi)e^{-\alpha_0r}
\end{equation}
where $\alpha_1+i\alpha_0$ is the pole of $\tilde G(k)$ in the upper half of the complex $k$-plane with
the smallest imaginary part and positive real part. The values of $\alpha_0$ and $\alpha_1$ depend significantly 
on the approximation for the entropy, and assuming the rather crude approximation in the MF theory,
we cannot expect quantitative agreement
with simulations. Still, a semiquantitative agreement is obtained for $\alpha_0$ that 
increases with $T$ in both theory and simulations.
Neglecting the effect of clustering and assuming the perfect gas approximation for the entropy, leads to much poorer 
agreement with simulations, especially for low $T$, where a large fraction of particles belongs to the clusters. 
In particular, assuming the perfect gas entropy for $T=0.11$, we obtain $\alpha_1=0.792$ and $\alpha_0=0.719$. 
$\alpha_1$ obtained in this MF theory agrees semiquantitatively with simulations, but 
decreases with $T$, in contrast to the simulation results.
 In table 1, we also present $2\pi/\ell$, where $\ell$ 
is the average distance between the objects defined simply by $\ell^3=\bar M/\rho$. It is interesting to compare
$\alpha_1$ and  $2\pi/\ell$, because $2\pi/\alpha_1$ describes the distance between the maxima of $g(r)$.
For high $T$, $\alpha_1^f$ and  $2\pi/\ell$ agree quite well, while for lower $T$, when the systems 
gets more ordered, the distance between the maxima of 
 $g(r)$ is smaller than the average distance between the objects.
 We conclude that fluctuations neglected in the MF theory influence 
the values of the parameters obtained from (\ref{G-1}), but 
the formula (\ref{gr}) is a good approximation already for separations larger 
than the period of the damped density oscillations.

\begin{figure}[h]
 \centering
 \includegraphics[scale=0.6]{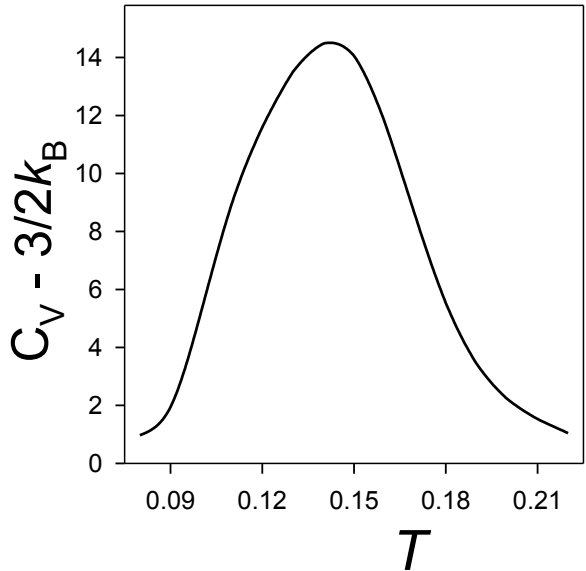}
\caption{$c_V-3k_B/2$, where  $c_V$ is the  specific heat, for $\rho=0.005$ as a function of temperature $T$.}
\label{fig_cv}
\end{figure}

Finally, we investigated the effect of clustering on the deviation of the specific heat form the perfect gas form.
As shown in Fig.\ref{fig_cv} for $\rho=0.005$, $c_V-3k_B/2$ takes a maximum for $T\approx 0.14$. 
For this temperature,
$p(1)\approx p(5)$ (see Fig.\ref{fig_his}). One can expect the largest fluctuations of the energy when
the probability that a particle is isolated or belongs to a cluster of the preferable size  is the same. 
The temperature $T_{c_V}$ corresponding to the maximum of $c_V$ is a borderline between dominating isolated particles (
$p(1)>p(5)$)
 and dominating clusters ($p(1)<p(5)$), above and below $T_{c_V}$, respectively.  The CCC temperature defined 
as the first appearance of nonmonotonic $p(M)$ upon decreasing $T$ is significantly higher.
Below $T_{CC}$ the clusters are formed, but
the monomers still dominate over the clusters until $T$ is decreased below $T_{c_V}$.

{\color{black}
The presence of a maximum in $c_V$  recalls the maxima associated with a thermodynamic transition. In Ref.\cite{godfrin:14:0} it is shown that there is a correspondence between the clustering region and the two-phase region of a reference fluid in which the long-range repulsion is absent. In the reference fluid, the  maximum in $c_V$ appears at the Widom line that is a continuation of the phase-coexistence line beyond the critical point in the supercritical phase. In our case, however, the density is much lower than the critical density. There is a common feature of the two cases, namely large fluctuations. While at
 the continuation of the phase transition line the fluctuations are of {\it a long range and a small amplitude} (large regions become a bit less or a bit more dense than the average density), in our case the fluctuations concern an aggregation of the monomers and the de-integration of the clusters,  i.e. have {\it a small range and a large amplitude}. 
The instantaneous changes of the numbers of the monomers and the optimal clusters lead to a significant fluctuations of the energy (the energy is much lower when a cluster is formed), thus to a large $c_V$. }

\section{structure near the wall}

In this section we study the effect of the attractive wall on the clustering in the near-surface layer, on 
the density profile $\rho(z)$, on the
adsorption $\Gamma(z_m)$ defined as 
\begin{equation}
\label{ads}
\Gamma(z_m)= \int_{z_L}^{z_m}(\rho(z)-\rho_g)dz,
\end{equation}
 and finally on the pair-distribution function 
in the layer adsorbed at the surface.

In the system with fixed number of particles, the gas density away from the left wall, $\rho_g$, depends on 
the strength of the wall-particle interactions (Eq.(\ref{VL})),
due to the adsorption on the attractive wall.
As a result,  at the equilibrium state the gas density far from the left wall 
was lower than the initial density $\rho_0=N/V$.
However, the difference was not high, and seldom exceeded $10\%$ relative value.  

\subsection{Critical cluster concentration}
\begin{figure}[h]
 \centering
 \includegraphics[scale=0.65]{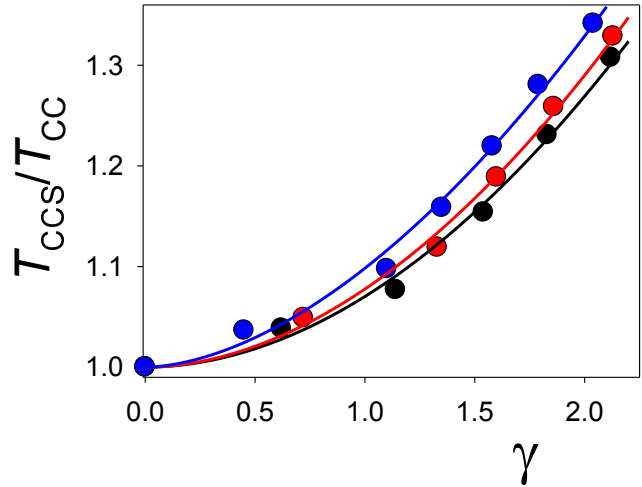}
\caption{  The ratio of the critical cluster concentration temperature at the  surface and in the bulk,
$T_{CCS}/T_{CC}$, for the same values of $\rho_0$. $\gamma$
is the strength of 
the particle-wall interaction. The symbols denote the simulation results, and the lines 
are fits to Eq.(\ref{fig_CCCS}).
Black,  red and  blue lines correspond to $\rho_0 = 0.0005, 0.0015,  0.005$ respectively.
 The gas density away from the surface, $\rho_g$, depends on $\gamma$. The effect is weak, however, 
 since the increase in $\gamma$ is accompanied by the increase in $T_{CCS}$.
The highest change in $\rho_g$ is from $\rho_g= 0.00465$ to $\rho_g= 0.0049$ along the curve for $\rho_0 = 0.005 $. }
\label{fig_CCCS}
\end{figure}

We first obtained the histograms for the clusters of the particles in the vicinity of the surface, i.e.
with $z<2.6$ (not shown). 
From these histograms,
we obtained the dependence of the temperature at the critical cluster concentration at the surface, $T_{CCS}$, on the 
attraction strength $\gamma$ (Fig.\ref{fig_CCCS}).

The simulation results can be
fitted quite well to the formula
\begin{equation}
 T_{CCS}/T_{CC}=1+B\gamma^{\beta_0}.
\end{equation}
 The fitting parameters for $\rho_0 = 0.0005, 0.0015,  0.005$
are $B= 0.07026,0.07765, 0.09833$ and $\beta_0=1.938,1.901, 1.742$ respectively.
The clustering occurs at significantly higher temperature near an adsorbing surface, and the increase of temperature 
is the larger the stronger the adsorption, and the larger the density in the bulk. One obvious reason for this
enhanced clustering is the larger density in the near-surface layer. This layer is a quasi-two dimensional system,
and this may influence the aggregation process as well. However, we cannot find an explanation for 
the power-law behavior with the exponent $\beta_0$ decreasing with increasing $\rho_0$.

\subsection{the density profile}

\begin{figure}[h]
 \centering
\includegraphics[scale=0.65]{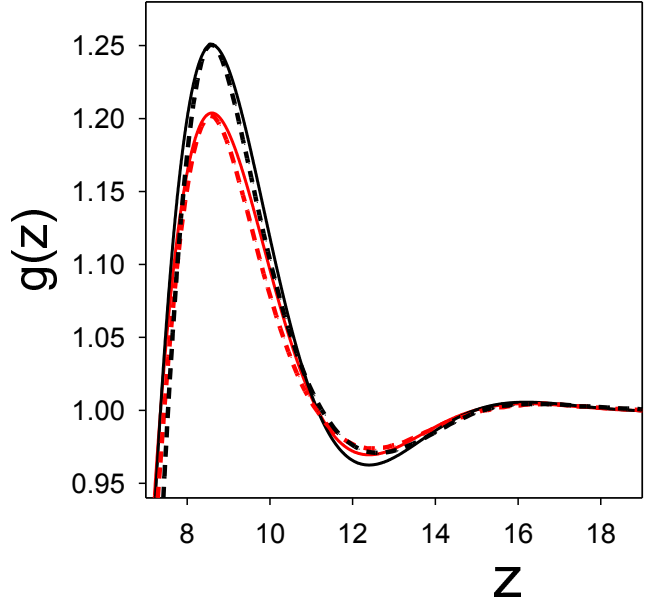}
\caption{The density profile for $\rho_0 = 0.005 $, $ T=0.13$. The dotted lines are the
simulation results, and the continuous lines are
the best fits to Eq.(\ref{g(z)}). The red  and black lines correspond to
$\gamma=0.5, 1.0$, respectively.  The $\alpha_i$ parameters in Eq. (\ref{g(z)})
are taken from the bulk gas simulation. From the fit we obtain $A=17.5,21.5$ and $\phi=0.2$ 
for $\gamma=0.5, 1.0$, respectively.
$z$ is in $\sigma$ units.
}
\label{fig_rho}
\end{figure}

The density 
averaged over the plane $(x,y)$ parallel to the wall as a function of the distance from the wall, $z$, should decay
in the same way as the pair distribution function. For $g(r)$ given by (\ref{g(r)}),
we expect that for large separations, 
\begin{equation}
\label{g(z)}
g(z):= \rho(z)/\rho_g=1+A \sin(\alpha_1 z+\phi)e^{-\alpha_0z}
\end{equation}
where $\alpha_1$ and $\alpha_0$ should take the same values 
 as in Eq.(\ref{gr}) for $g(r)$ at the same thermodynamic conditions. 
The remaining parameters depend on $\gamma$. The simulation results and the fit to Eq.(\ref{g(z)}) are shown
in Fig.\ref{fig_rho} for $z>6$. The agreement is satisfactory.

\begin{figure}[h]
 \centering
 \includegraphics[scale=0.55]{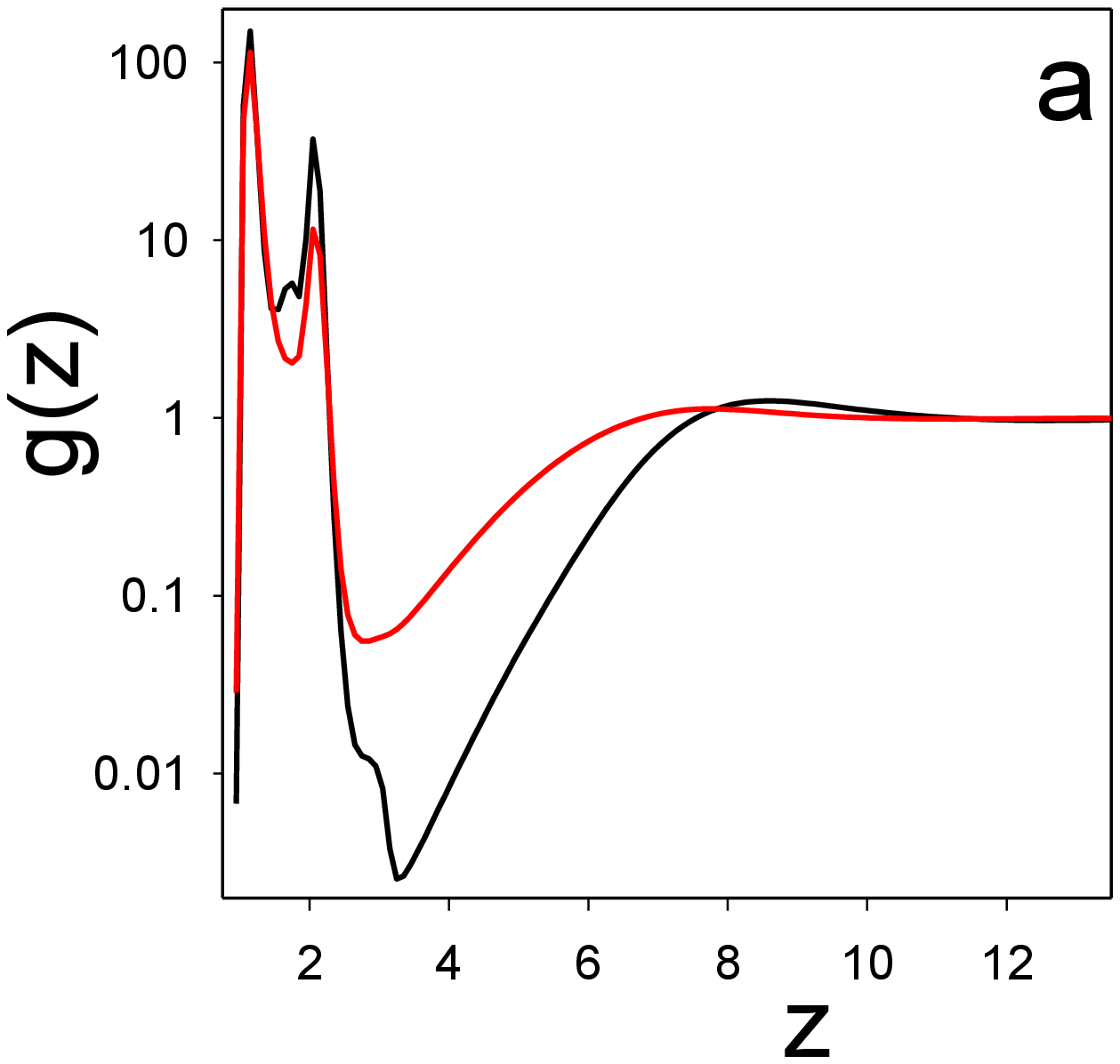}
 \includegraphics[scale=0.55]{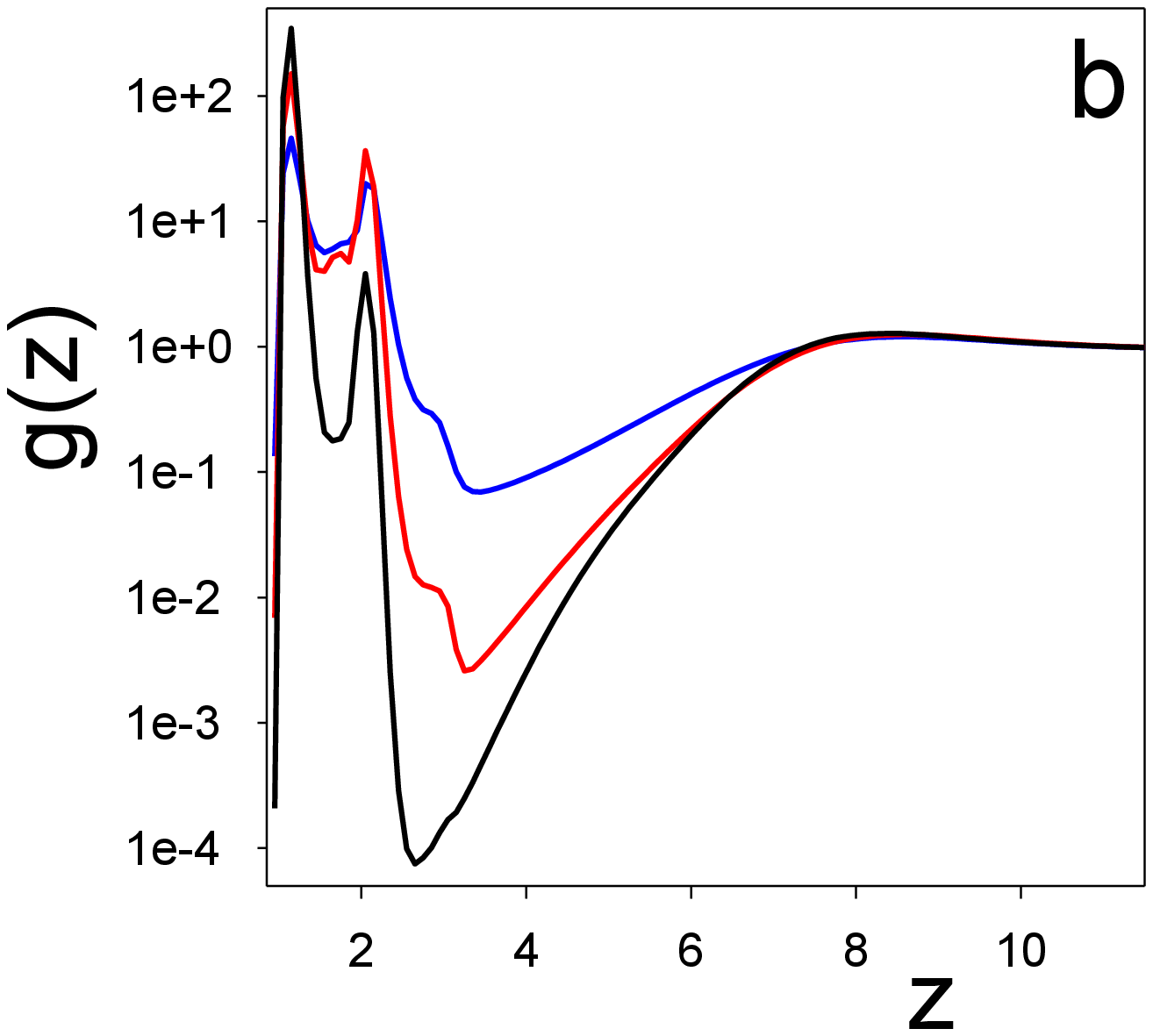}
\caption{The density profile for $\rho_0 = 0.005 $; (a): $\gamma=1$, and  $ T=0.13,0.17$
for the upper (red) and lower (black) line, respectively, (b): $T=0.13$ and  
from the bottom to the top lines (black, red, blue)  $\gamma=1.5,1,0.5$.
$z$ is in $\sigma$ units.
}
\label{fig_rho1}
\end{figure}

 Much more interesting is the density profile close to the surface. 
In Fig.\ref{fig_rho1} we present $g(z)$ for different $T$ and $\gamma$.
The two maxima 
corresponding to the first and the second layer of the
particles adsorbed at the surface, is followed by a deep and wide minimum that extends up to $z\approx 6$.
Note the  very small average density
for $z\approx 3$. The ratio of the density for $z\approx 3$ and the density in the gas away from the surface
decreases significantly with increasing wall-particle attraction, and becomes 
as small as $10^{-4}$ for $\gamma=1.5$. The essentially empty region rather close to the adsorbing surface
results from the formation of a layer consisting of particles that repel each other at distances
larger than $r \approx 2$.
Accumulated repulsion from the adsorbed particles exceeds the wall-particle attraction for $z\ge 2.5$,
and leads to the depletion
of the particles beyond the bilayer formed at the surface. 

\begin{figure}[h]
 \centering
\includegraphics[scale=0.7]{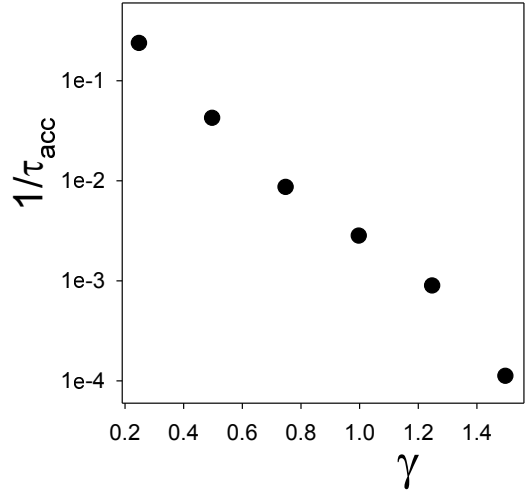}
\caption{The inverse accessibility time $\tau_{acc}$ defined in Eq.(\ref{tau}) for $T=0.13$
as a function of the strength of the wall-particle attraction. 
}
\label{fig_tau}
\end{figure}

The repulsion from the surface for $2.5\le z\le 6$ is  stronger when more particles are adsorbed in the two near-surface 
layers. The strong repulsion barrier prevents from further adsorption of the particles 
that must overcome this barrier in order to 
enter the region close enough to the surface, where the attraction dominates (provided that the distance from
the adsorbed particles is either smaller than $r\approx 2$ or larger than the range of the repulsion).
The formation of the repulsive zone beyond the adsorbed bi- or monolayer of particles,
has a significant effect on the dynamics of the adsorption process. 
Inspired by the Smoluchowski-equation based theory of chemical reactions 
\cite{pedersen:81:0,szabo:89:0}, and noting that $g(z)^{-1}$ can be a 
measure of the repulsive barrier,
we introduce the ``accessibility time'' by
\begin{equation}
\label{tau}
 \tau_{acc}=\int_{z_1}^{z_2}g(z)^{-1}dz,
\end{equation}
where for the lower and the upper boundary we assume the position of the second maximum of $g(z)$,
and $z_2=z_1+\pi/\alpha_1$, respectively, where $\pi/\alpha_1$ is half the period of 
the density wave (see (Eq.(\ref{g(z)})). 
In Fig.\ref{fig_tau}, $ 1/\tau_{acc}$ is shown for $T=0.13$ as a function of $\gamma$.

The plot suggests approximately exponential increase of the accessibility time with $\gamma$. At first sight,
it seems counter-intuitive that the stronger the attraction to the surface, the larger time is required  to 
approach it.  
This paradox is a direct consequence of the competing interactions and formation of the layer of particles 
that ``screen'' the attraction to the wall by their repulsive interactions. In MD simulations we indeed see
a significant
slowing down of the evolution, which makes it difficult to reach the equilibrium state.  
An anomalous dynamics in the SALR system was recently observed in Ref.~\cite{zhuang:17:0,bergman:19:0}.

\subsection{The adsorption}

\begin{figure}[h]
 \centering
 \includegraphics[scale=0.6]{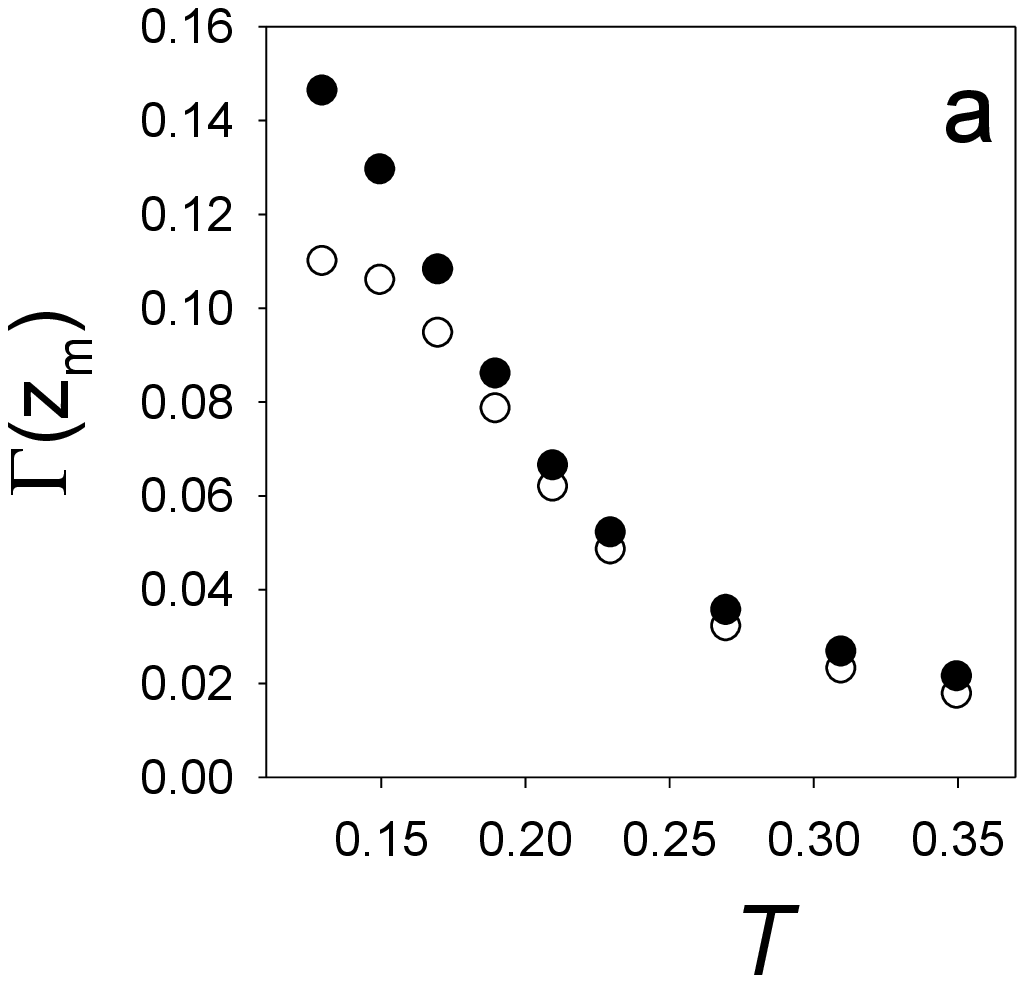}
  \includegraphics[scale=0.6]{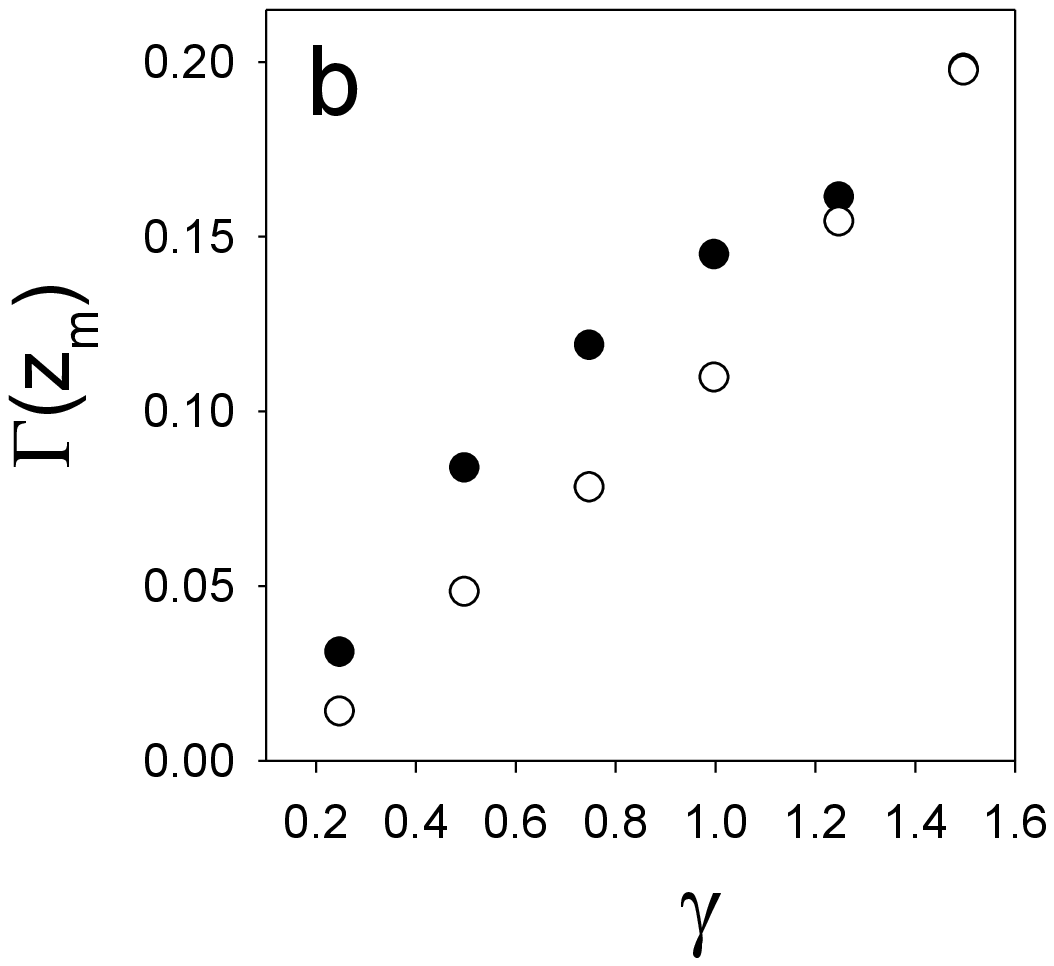}
\caption{The adsorption (\ref{ads}) for  $\rho_0 = 0.005 $ and $z_m=1.5, 2.6$ 
(open and filled symbols, respectively).
(a): as a function of $T$ for $\gamma=1$.  (b): as a function of $\gamma$ for $T=0.13$.
}
\label{fig_ads}
\end{figure}

The adsorption $\Gamma(z_m)$ (Eq.\ref{ads}), calculated for $z_m=1.5, 2.6$ is shown in Fig. \ref{fig_ads}
as a function of $T$ and $\gamma$.

For high $T$ or for strong attraction,  $\Gamma(1.5)\approx \Gamma(2.6)$,
indicating that the particles occupy
practically only the first layer. In particular, if $\gamma=1$, the adsorption occurs only in the first layer 
for $T>0.18\approx  T_{CCS}$, i.e. when isolated particles are present near the surface.
Below $T_{CCS}$, when clusters appear at the surface, some particles
forming the cluster are located in the second layer at $z\approx 2$. This is consistent with the second maximum 
of $g(z)$ (Fig.\ref{fig_rho1}a).
On the other hand, the strong attraction to the surface can overcome the particle-particle repulsion,
leading to deformation of the clusters. In particular, at $T=0.13$ the clusters become essentially flat for 
$\gamma>1.2$, consistent with  Fig.\ref{fig_ads}b ($\Gamma(1.5)\approx \Gamma(2.6)$) and Fig.\ref{fig_rho1}b (the second peak of $g(z)$ is
about two orders of magnitude smaller than the first one).

\subsection{structure of the adsorbed layer of particles}

\begin{figure}[h]
 \centering
 \includegraphics[scale=0.55]{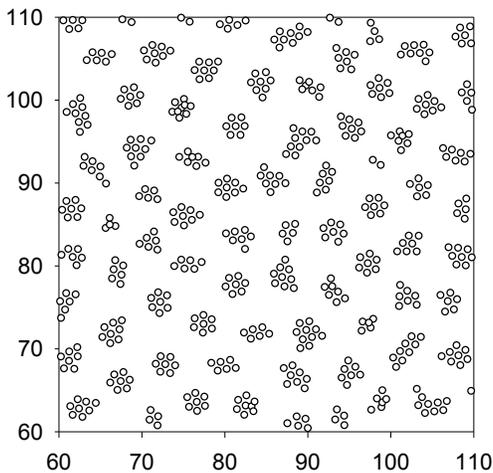}
\caption{A projection of a representative configuration  of the particles  adsorbed at the surface
for  $\rho_0=0.005$, $T=0.13$, $\gamma=1.5$ and the adsorption  $\Gamma(2.6)\approx 0.2$.
The diameter of the shown circles is $r_{min}= 1.139$.
}
\label{fig_snap}
\end{figure}

\begin{figure}[h]
 \centering
 \includegraphics[scale=0.6]{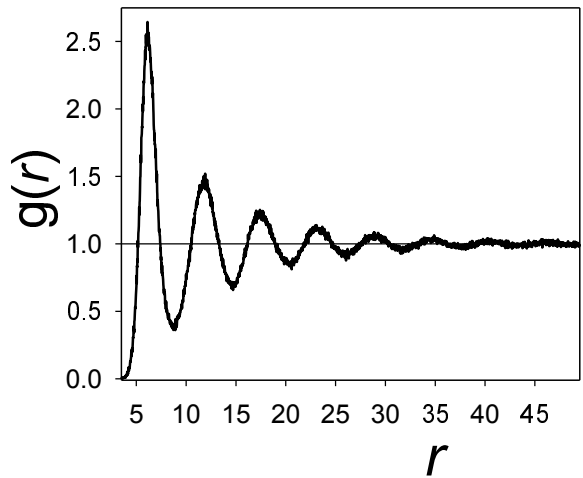}
\caption{
 2D pair distribution function for centers of mass for $M\ge 4$ for $\rho_0=0.005$, $T=0.13$
and $\gamma=1.5$. The adsorption for this case is $\Gamma(2.6)\approx 0.2$, 
and a representative configuration is shown in Fig.\ref{fig_snap}.
}
\label{fig_snap2}
\end{figure}

A snapshot of the particles adsorbed at the strongly attracting surface with $\gamma=1.5$ for $\rho_0=0.005$ and 
$T=0.13$ is shown in Fig.\ref{fig_snap}. 
Most of the particles are aggregated into clusters with $M>4$ when attraction to the surface is that strong. 
Many clusters
have a form of a 7-particle ``flower'' with six particles surrounding the central one, but there exist 
also larger clusters, with one or two additional particles attached to the flower. 
Another typical shape is a short piece of a straight bilayer. Irregular shapes  appear too.
The separation between the 
neighboring clusters is roughly the same, and locally a hexagonal distribution of the clusters can be observed. 
The long-range periodic arrangement of the clusters could not be observed, however. 
The pair distribution function for the centers of 
mass of the clusters in the $(x,y)$ plane exhibits an oscillatory decay with a large decay length 
(Fig.\ref{fig_snap2}).

\begin{figure}[h]
 \centering
\includegraphics[scale=0.55]{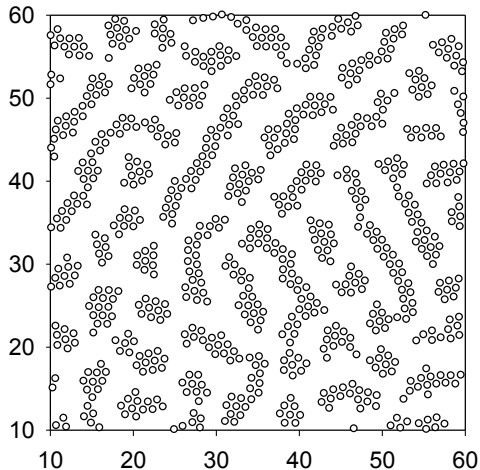}
\caption{A projection of the representative configuration  of the particles  adsorbed at the surface
for  $\rho_0=0.00676$, $T=0.15$, $\gamma=2.5$ and the adsorption $\Gamma(2.6)\approx 0.315$.
The diameter of the shown circles is $r_{min}= 1.139$.}
\label{fig_snap3}
\end{figure}

 In Fig.\ref{fig_snap3} we show a configuration at the  surface with stronger wall-particle attraction ($\gamma=2.5$). 
 In this case, stripes of different length form an isotropic labyrinth.
In equilibrium we expect stripes  too, 
but it is not clear wheather the isotropic or anisotropic structure with preferred orientation 
of stripes, found for a similar 2D model in Ref.\cite{almarza:14:0}, will occur. 
Due to the exponentially growing time scale of evolution near the adsorbing surface, it 
is difficult to reach the equilibrium.

\section{summary and conclusions}

The purpose of our study was determination of general features of adsorption phenomena in dilute systems
 with particles self-assembling into small clusters. Our MD simulations were performed for a generic model
 with the SALR potential (\ref{u(r)}), for various strengths of attraction between the particles and a flat surface.
 
 In the first step we focused on the aggregation in the bulk. We determined the temperature at the 
 critical cluster concentration for three different densities according to the definition of the CCC introduced
 in Ref.\cite{santos:17:0}. At the CCC, the probability of finding a particle in the cluster composed of $M$ 
 particles becomes nonmonotonic. Our histograms and the specific heat (Figs.\ref{fig_his} and \ref{fig_cv}), 
 indicate that one can
 introduce another structural line determined by the maximum of $c_V(T)$. At this line, the probability of finding an
 isolated particle and a particle inside the cluster of the most probable size are equal. Thus, 
 the maximum of  $c_V(T)$ represents a crossover from the monomer dominated to the cluster dominated
 regime in the gas.
 
 We have calculated the pair distribution function $g(r)$. The simulation results fit quite well
 the simple formula (\ref{gr}) for $r$ larger than the period of the oscillatory decay of $g(r)$. 
 
 Near an attractive surface the density of the particles increases, and one can expect an increase 
 of the critical cluster  concentration temperature in the near-surface layer.
 Indeed, we found a significant increase of $T_{CCS}$ with increasing strength of the wall-particle attraction. 
 Simulations indicate that  $T_{CCS}(\gamma)/T_{CC}-1\propto \gamma^{\beta_0}$ with the exponent $\beta_0$ 
 depending on the gas density. Further studies are required  to explain the origin of this behavior.
 
 The calculated amount of particles adsorbed in the monolayer and in the bilayer at the surface  allows us to 
 follow the scenario of the process of adsorption. When the wall-particle attraction is moderate,  and
the gas consists of the isolated particles ($T>T_{CCS}$), then the
 particles are adsorbed in the monolayer at the surface.
 Below the CCC line, the clusters get adsorbed at the surface, and as they are 3D objects, some part of the cluster 
 occupies the second layer at  the surface. However, when $\gamma$ increases, the clusters become flattened, and 
 the excess of density in the second layer decreases. The structure in the first monolayer is shown in Fig.\ref{fig_snap}.
We can see that despite the strong wall-particle attraction, a significant fraction of the surface area is
not covered by the particles, because they repel each other at large separations. 
 
 Perhaps the most interesting result that distinguishes strongly the SALR system from simple fluids, is the 
 formation of a depletion layer just outside the bilayer adsorbed at the surface. The stronger the wall-particle attraction, 
 the smaller the density  in the depletion zone. For strong attraction, this depletion  layer is essentially empty.
 This is because the adsorbed bilayer ``screens'' the attraction of the surface, and the accumulated repulsion 
 from the adsorbed particles forms a large repulsive barrier. 
 We can thus observe an effective repulsion from the attractive surface. The barrier grows in the
 process of adsorption,
 and this leads to slowing down of the adsorption - the effect is the 
 stronger, the larger the attraction to the surface.
 
 The effect described above is not specific to the particular model chosen for the simulations, and follows from the presence of the repulsive tail 
 in the interactions. 
 Our results indicate that the adsorption process in a self-assembling system differs significantly from the process
 in simple fluids.
 Large strengths of the wall-particle attraction lead to strong effective repulsion beyond the layer adsorbed at the 
 surface. Based on our results, one can expect very nontrivial dynamics of formation of ordered patterns,
 as has been already observed in Ref.~\cite{zhuang:17:0,bergman:19:0} for the SALR systems in bulk.
 
{\color{black} It would be interesting to verify our predictions experimentally. An experimental system with the SALR interactions leading to formation of small clusters is for  example a much studied dilute solution of lysozyme molecules in deionized water. An attractive surface in this case is for example a weakly charged electrode with an opposite sign, although electrostatic effect may play some role in this case. Another example is provided by weakly charged nanoparticles, with the short-range attraction induced by some kind of small depletion agents present in the solvent, near a surface covered by the same material as the surface of the nanoparticles. }
 
\section{Acknowledgements}
ML  would like to thank prof. Vikhernko for discussions and  hospitality during his stay at Belarusian State Technical University.
AC is grateful to prof. Zarragoicoechea, Dr. Meyra and Dr. de Virgiliis for discussions, comments on the manuscript and  hospitality at the National University of La Plata.
 This project has received funding from the European Union Horizon 2020 research 
and innovation programme under the Marie
Sk\l{}odowska-Curie grant agreement No 734276 (CONIN).
An additional support in the years 2017-2018  has been granted  for the CONIN project by the Polish Ministry
of Science and Higher Education. 
Financial support from the National Science Center under grant No. 2015/19/B/ST3/03122 is also acknowledged.


\end{document}